\documentclass[aps, pra, twocolumn, notitlepage, nofootinbib, longbibliography]{revtex4-1}

\usepackage{graphicx,amsmath,pstricks,float,subcaption,mathtools}

\begin{document}

\title{Implications of a search for intergalactic civilizations on prior estimates of human survival and travel speed}

\author{S. Jay Olson}
 \email{stephanolson@boisestate.edu}
 \affiliation{Department of Physics, Boise State University, Boise, Idaho 83725, USA}
\author{Toby Ord}
 \email{toby.ord@philosophy.ox.ac.uk} \thanks{This research was supported by the European Research Council (ERC) under the European Union’s Horizon 2020 research and innovation programme (grant agreement No 669751).}
 \affiliation{Future of Humanity Institute, University of Oxford, Oxford, UK}

\date{\today}

\begin{abstract}
We present a model where some proportion of extraterrestrial civilizations expand uniformly over time to reach a cosmological scale. We then ask what humanity could infer if a sky survey were to find zero, one, or more such civilizations. We show how the results of this survey, in combination with an approach to anthropics called the Self Indication Assumption (SIA), would shift any prior estimates of two quantities: 1) The chance that a technological civilization like ours survives to embark on such expansion, and 2) the maximum feasible speed at which it could expand. The SIA gives pessimistic estimates for both, but survey results (even null results) can reverse some of the effect. The SIA-induced shift in expectations is strong enough to be regarded as a falsifiable technological prediction, and a test of the SIA in cosmological reasoning. 
\end{abstract}

\maketitle

\section{Introduction}
The standard motive for studying the possible behavior and technology of intelligent extraterrestrial life is to generate search targets for SETI. This has been true for the development of ``expanding cosmological civilizations"~\cite{olson2018, olson2018b} (aka ``aggressively expanding civilizations,"~\cite{olson2014} ``loud aliens,"~\cite{hanson2021} etc.) --- a behavioral endpoint for expansionistic civilizations in the universe, based on conservative physics assumptions, but ambitious technologies. The resulting cosmology is precise enough that one can use anthropic reasoning to estimate the appearance rate of such expansionistic civilizations~\cite{olson2017a, olson2020}, and their visible geometry~\cite{olson2016}, as guidelines for future searches --- a form of extragalactic SETI. 

Here, we take a step in a different direction, using the same basic model of expanding civilizations. We use anthropic reasoning to estimate two additional parameters in the cosmology --- the maximum practical expansion speed in the universe, $v$, and the probability that any given human-stage civilization will eventually become expansionistic, $q$. These estimates have a far more personal meaning than most in cosmology. They can be interpreted as predicting aspects of humanity's own near-term future.

What we show is how one's prior estimates of $q$ and $v$ change when adopting an anthropic school (the Self-Indication Assumption~\cite{bostrom2002}, or SIA) combined with the world model of expanding cosmological civilizations. If one's prior estimates are given as probability density functions $P(q)$ and $P(v)$, adopting this cosmology re-weights $P(q)$ and $P(v)$ to lower than expected values of $q$ and \emph{much} lower than expected values of $v$. Specifically, $P(q) \rightarrow \frac{P(q)}{q}$ and $P(v) \rightarrow \frac{P(v)}{v^3}$, properly normalized.

This re-weighting of expectations can be extremely powerful, particularly in the case of $v$. For any plausible prior estimate $P(v)$ that covers a substantial range of $v$ (e.g. giving some weight to the possibility of both $v = .1 \, c$ and $v = .9 \, c$), the update will place enormous weight on the lowest values of $v$. This is strong enough to be regarded as a near-term technological prediction, with reference to projects in current development like Breakthrough Starshot~\cite{merali2016}, which target a spacecraft speed of $v = .2 \, c$.

In a further step, we show how estimates for $q$ and $v$ can be changed again. For $v$, this occurs when a cosmological sky survey is performed, searching for expanding civilizations. If one or more are detected, $v$ could be inferred from their visible geometry~\cite{olson2016}. But even if $n=0$ civilizations are detected, $P(v)$ is shifted back in the direction of one's original prior estimate --- the extent to which it is shifted back will depend on the certainty of the search result.

An analogous update to $P(q)$ occurs when humanity develops the technology to embark on cosmic expansion. The act of successfully navigating ``late filters''~\cite{hanson1998a} pushes estimates of $q$ back in the direction of the original prior estimate. 

We organize the paper as follows: Section II discusses the inputs to our analysis, reviewing the basic assumptions about expanding cosmological civilizations, assumptions about our prior pdf's, and the Self-Indication Assumption. Section III derives our initial predictions based on adopting the cosmology, while section IV derives the update from search results. Section V is devoted to a discussion of $q$ and increased existential risk for humanity. Section VI reflects on the meaning and power of these results, discussing the implied predictions and test of the SIA, and contains our concluding remarks.

\section{Inputs}
The basic ingredients for this kind of analysis are a \emph{world model} (depending on a set of parameters), a set of \emph{prior assumptions} about the world (manifested by a probability density function over the parameters), and an \emph{anthropic school} that describes how we should account for observer-selection effects.

We review these here. The ``world model'' is the cosmology of expanding cosmological civilizations and the ``anthropic school'' for this analysis is the Self-Indication Assumption.

\subsection{Expanding cosmological civilizations}

If a small set of technologies are practical for an advanced civilization, then a form of extreme expansionist behavior becomes possible. The essential technology is fast spacecraft that can travel between galaxies~\cite{fogg1988, armstrong2013}, reproduce themselves from available resources~\cite{freitas1980}, and carry out general instructions. If this technology is physically attainable, then a single actor can release a wave of intergalactic settlement that expands (over cosmological time) in all directions at speed $v$, fully saturating/harnessing every galaxy along the way. The cost of such a cosmological project is just the cost of launching the first spacecraft. The reward is control of perhaps thousands of superclusters of galaxies.

If the appearance of such expansionistic civilizations happens rarely and at random in the universe, this behavior results in a clean geometry at the cosmological scale, where the universe is homogeneous --- in fact the geometry is almost identical to nucleation and bubble growth in a cosmological phase transition~\cite{guth1981,olson2014}. The following assumptions give rise to the basic geometrical picture we use:

\begin{enumerate}
	\item The background cosmology is standard $\Lambda$CDM --- a spatially flat Friedmann-Robertson-Walker solution\footnote{We assume the following solution:  $\Omega_{\Lambda 0}=.692$, $\Omega_{r0}=9 \times 10^{-5}$, $\Omega_{m0}=1-\Omega_{r0} -\Omega_{\Lambda 0}$, $H_0 =.069 \, Gyr^{-1}$.}, in which we have appeared at cosmic time $t_0 = 13.8$ Gyr. We denote the scale factor as $a(t)$.
	\item Appearance events are a spatially random process in the universe (a Poisson point process), with an appearance rate denoted as $f(t)$, with units of ``appearances per $Gly^3$ of co-moving coordinate volume per $Gyr$ of cosmic time.''
	\item Each appearance event is followed by expansion in all directions at constant speed $v$ in the co-moving frame of reference, resulting in a growing domain of fully occupied galaxies\footnote{We take units in which $c=1$, and so for the rest of this paper, we express $v$ as a number strictly less than one.}. The net expansion speed of the domain will be close to the speed of the intergalactic spacecraft, since the timescale for travel between galaxies dominates all other timescales in the process (e.g. reproduction, launch, deceleration, etc.). This type of galaxy-to-galaxy expansion can easily be modeled with a simulated distribution of galaxies, and it results in a growing spherical domain with a sharp boundary, above the homogeneity scale~\cite{olson2018}.
	\item We assume that, once initiated, the process of cosmic expansion does not stop unless constrained by the presence of other expanding domains, or the causally isolating effect of cosmic acceleration in the distant future~\cite{ord2021edges} (strictly limiting all domains to a finite maximum size, depending on the time of their appearance).   
\end{enumerate}

We do not assume that all technological civilizations initiate such expansion. We introduce a parameter $q$ to represent the fraction of human-stage civilizations that go on to cosmic expansion. Failure to expand could come from any cause, from voluntary abandonment of technology~\cite{joy2000} to ``late filter" extinction~\cite{hanson1998a}. The value of $q$ only tells us the fraction of human-stage civilizations that result in cosmic expansion, or in other words, $q$ is the probability that a random human-stage civilization goes on to expansion.

The cosmic appearance rate of expanding civilizations, $f(t)$, can be expressed as $f(t) = \alpha \, F(t)$, where $\alpha$ is an uncertain parameter, setting the scale of the appearance rate, and $F(t)$ is a model for the relative time dependence of the appearance rate of Earthlike planets (normalized to a maximum value of unity), with a lag of several $Gyr$ to account for the slow process of biological evolution. While the $F(t)$ we use (displayed in figure 1) should not be too controversial~\cite{loeb2016}\footnote{Substantially different models for $F(t)$ have recently been proposed~\cite{hanson2021} in the context of a ``hard steps'' model of evolution that amplifies the appearance rate by a factor of $t^n$, for some model-dependent $n$.}, $\alpha$ sets the scale and is radically uncertain over many orders of magnitude.

\begin{figure}
	\centering
	\includegraphics[width=0.45\textwidth]{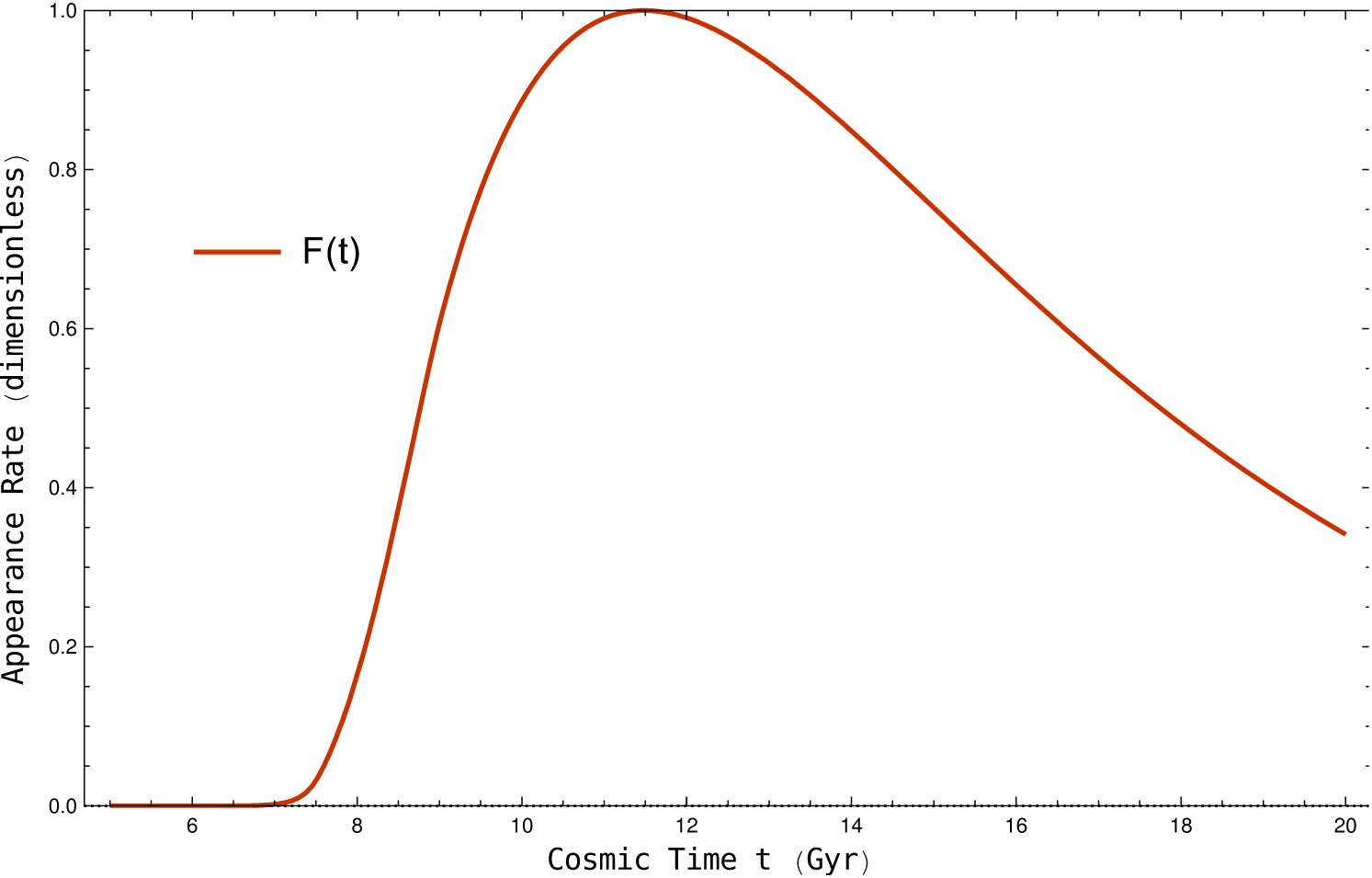}
	\caption{A model for the cosmic time-dependence of the relative appearance rate of advanced life, $F(t)$, which is effectively a model of the planet formation rate of the cosmos, with a lag of several Gyr to allow for the evolution of intelligent life.}
\end{figure}

We can further relate $\alpha$ to human-stage civilizations by expressing $\alpha = \gamma \, q$, where $\gamma \, F(t)$ represents the cosmic appearance rate of human-stage civilizations. That is, the appearance rate of expanding civilizations is the appearance rate of human-stage civilizations, times the probability for a human-stage civilization to expand. Thus, we now have $f(t) = \gamma \, q \, F(t)$.

When an ambitious civilization appears at cosmic time $t'$ and initiates cosmic expansion (an appearance event), it will occupy a growing sphere with co-moving coordinate volume given by:

\begin{eqnarray}
v^3 \, V(t',t) = v^3 \,\frac{4 \pi}{3} \, \left( \int_{t'}^t \frac{1}{a(t'')} dt'' \right)^3.
\end{eqnarray}

Here, $V(t',t)$ is the co-moving coordinate volume of a sphere of light at time $t$ that was emitted at $t'$, and thus multiplying by $v^3$ gives the coordinate volume of an expanding civilization\footnote{In an expanding universe, a coasting spacecraft will eventually be slowed relative to the co-moving frame~\cite{tipler1999}, over cosmic time. However, since the growth of a civilization will presumably generated by galaxy-to-galaxy hops, with the speed reset at each galaxy, we expect the net expansion speed $v$ to be constant and close to the spacecraft speed at the time of launch~\cite{olson2018}.}. We assume that all ambitious civilizations expand with a similar speed (the maximum practical speed), and thus we call $v$ the dominant expansion velocity, treated as another uncertain parameter.

With the assumption that ambitious civilizations appear at random and independently in the cosmos, the fraction of all space that is \emph{not} occupied by them is denoted by $g_{\alpha, v}(t)$, given by~\cite{olson2014}:

\begin{eqnarray}
g_{\alpha,v} (t) = e^{- \alpha v^{3} s(t)}.
\end{eqnarray}

Here, $s(t) = \int_{0}^{t} F(t') \, V(t',t_0) \, dt'$, i.e. the integral over all space within our past light cone, weighted by $F(t)$ --- for our cosmological parameters and model for $F(t)$, we get $s(t_0) \approx 1268 \, Gly^3 \, Gyr$.

The average/expected number of expanding civilizations on our past light cone (and thus in principle visible to us) is denoted $E_{\alpha, v}$, given by~\cite{olson2017a, olson2016}:

\begin{eqnarray}
E_{\alpha,v}=\alpha (1 - v^3) s(t_0).
\end{eqnarray}

A survey that covers only a fraction $frac$ of the sky, or for any other reason expects to detect only a fraction $frac$ of civilizations on our past light cone would thus expect to detect, on average, $ frac \, E_{\alpha,v}$.

A cosmological expansion scenario is set by the parameter pair $\{\alpha, v \}$, and previous calculations have focused on estimating $\alpha$, assuming a given value of $v$. Since we are interested here in predictions that are relevant to humanity's future, we will work with the parameterization $\{\gamma, q, v \}$. All equations above expressed in terms of $\alpha$ can be obtained in the new parametrization with the replacement $\alpha \rightarrow \gamma \, q $.   

\subsection{Prior probability}

Before any anthropic calculations, we should specify our prior assumptions about the parameters $\{\gamma, q, v \}$. We will assume an independent prior pdf for each, so that $P(\gamma, q, v) = P(\gamma) \, P(q) \, P(v)$.

Our results in the following sections will concern how to \emph{update} $P(q)$ and $P(v)$ using our anthropic model, for any prior one might adopt. Thus, we keep these general, for now. But we do specify a particular form for $P(\gamma)$ --- it is thus an important ingredient of our final results.

The rate of appearance of human-stage civilizations in the universe is radically uncertain over many orders of magnitude. In a prior sense, $\gamma$ appearing in any order of magnitude from $10^{10}$ to $10^{-100}$ (appearances per $Gly^3$ per $Gyr$) is equally plausible, but even this could be criticized as too narrow, with other estimates in the literature spanning as many as $10^{122}$ orders of magnitude~\cite{lacki2016}.

This degree of prior uncertainty suggests beginning with a log-uniform prior~\cite{tegmark2014}, i.e.
$P(\gamma) = \frac{1}{\ln(\gamma_{max}/\gamma_{min})} \frac{1}{\gamma}$. Indeed log-uniform behavior can be seen to emerge in a related context, when compounding many sources of modest uncertainty to make estimates with the Drake equation~\cite{sandberg2018}. The endpoints of our prior, $\gamma_{min}$ and $\gamma_{max}$, should be separated by many orders of magnitude. We will find that if $\gamma_{min}$ is sufficiently small (below about $10^{-3}$), and $\gamma_{max}$ is sufficiently high (above about $10^4$), pushing the endpoints further out will not effect our final results.   

\subsection{Self-indication assumption}

The classic statement of the Self-Indication Assumption is that \emph{we should reason as though we are randomly selected from the set of all possible observers}~\cite{bostrom2002}. In the context of the world-model above, this means that if we have some prior pdf $P(\gamma, q, v)$ over the parameters (before anthropic arguments), the SIA implies that we should re-weight the prior, proportional to the total number of observers appearing in each parameter-universe (per unit co-moving volume, over all time~\cite{bostrom2003b}), $N_{obs}(\gamma, q, v )$, giving a ``SIA prior" proportional to $N_{obs}(\gamma, q, v ) P(\gamma, q, v)$, and re-normalized.

That is, the ``SIA prior" says we are more likely to live in a parameter-universe $\{\gamma, q, v \}$ that contains many observers, before we incorporate any further knowledge. It is probably hopeless to find a good model for $N_{obs}(\gamma, q, v )$ --- the total number of observers would presumably be dominated by members of vast and long-lived civilizations in the distant cosmic future, and thorny questions of the ``observer reference class" are inevitable. 

But, the story is not complete --- we should include our particular anthropic information. That is, we humans are not just \emph{any} observers, we are observers from a human-stage civilization that appeared at cosmic time $t_0 = 13.8 \, Gyr$, in an otherwise empty corner of the cosmos (devoid of expanding civilizations).

Incorporating this anthropic information will take the form of a Bayesian ``update" to the SIA prior --- for now we keep the information in the update general, simply denoting it $P(\gamma, q, v | \, \textnormal{info} )$. According to Bayes' theorem, this is given by
\begin{widetext}
\begin{eqnarray}
P(\gamma, q, v | \, \textnormal{info} ) = \frac{L(\gamma, q, v | \textnormal{info}) \, N_{obs}(\gamma, q, v ) \, P(\gamma, q, v)}{\int \int \int L(\gamma', q', v' | \textnormal{info}) \, N_{obs}(\gamma', q', v' ) \, P(\gamma', q', v') \, d \gamma'  dq' dv' }
\end{eqnarray}
where $L(\gamma, q, v | \textnormal{info})$ is the likelihood function, i.e. the probability that a random observer in world $\{\gamma, q, v \}$ sees ``info." This is given by $L(\gamma, q, v | \textnormal{info}) = N_{obs}( \textnormal{info} | \gamma, q, v ) / N_{obs}(\gamma, q, v )$.

Substituting this form of $L(\gamma, q, v | \textnormal{info})$ into the previous equation, the factors of $N_{obs}(\gamma, q, v )$ cancel out, giving:

\begin{eqnarray}
P(\gamma, q, v | \, \textnormal{info} ) = \frac{N_{obs}( \textnormal{info}|\gamma,q,v ) \, P(\gamma, q, v)}{\int \int \int N_{obs}( \textnormal{info}|\gamma',q',v')  P(\gamma', q', v') \,  d \gamma' dq' dv' }.
\end{eqnarray} 
\end{widetext}

Unlike the SIA prior, which depended on the unknowable $N_{obs}(\gamma, q, v )$, this expression only requires a model for $N_{obs}( \textnormal{info}|\gamma,q,v )$, the number of observers with anthropic information like ours. This is something we \emph{can} hope to model --- indeed, we have already introduced the world model to do it.

From equation 5, one can see the SIA now says that we should favor parameter-worlds containing many observers \emph{with anthropic info identical to our own}. It no longer blindly favors more total observers throughout all of spacetime, just more like us. In our case, the ``anthropic info" is that we are members of a human-stage civilization, at cosmic time of arrival $t_0 = 13.8$ Gyr, in a region of the universe that is \emph{not} saturated by expansionistic life.

In the world model of the previous section, the number of human-stage civilizations in parameter-world $\{\gamma, q, v \}$ who see this anthropic info (``info" denoted now by $t_0$) is proportional to: 
\begin{eqnarray}
N_{obs}( t_0 | \gamma, q , v ) \propto \gamma \, g_{\gamma,q,v}(t_0).
\end{eqnarray}

That is, the number of human-stage observers in this universe is proportional to their appearance rate $\gamma$ in empty space, times the fraction of space (at the current time) that has not been saturated with ambitious civilizations.

This gives us the machinery to calculate posterior probabilities, i.e. the SIA updates to any chosen prior.

\section{Predictions prior to cosmological searches for life}

As expressed above, $N_{obs}( t_0 | \gamma, q , v ) \propto \gamma \, g_{\gamma,q,v}(t_0)$, i.e. the cosmic number (density) of human-stage observers who note they have spontaneously appeared at cosmic time $t_0$ in unoccupied space, is proportional to the fraction of unoccupied space at the present time, $g_{\gamma,q,v}(t_0)$, times the appearance rate of human-stage civilizations appearing in unoccupied space, $\gamma$.

Thus, by equations 5 and 6, the SIA posterior over $\{\gamma, q, v \}$, according to this world-model, is:

\begin{widetext}
\begin{eqnarray}
P(\gamma, q, v | t_0) = \frac{\gamma \, g_{\gamma,q,v}(t_0) \, P(\gamma) \, P(q) \, P(v)}{\int_{\gamma_{min}}^{\gamma_{max}} \int_{q_{min}}^{q_{max}} \int_{v_{min}}^{v_{max}}  \gamma' \, g_{\gamma',q',v'}(t_0) P(\gamma')  P(q')  P(v') \,  d\gamma' \, dq' \, dv' }.
\end{eqnarray}

Substituting our model $g_{\gamma, q, v}(t_0) =  e^{- \gamma \, q \, v^{3} \, s(t_0)}$ (equation 2) and the log-uniform prior for $P(\gamma)$, then integrating over $\gamma'$, this becomes:

\begin{eqnarray}
P(\gamma, q, v | t_0) = \frac{s(t_0) \, e^{- \gamma q v^3 s(t_0) }  \, P(q)  \, P(v)}{    \int_{q_{min}}^{q_{max}} \int_{v_{min}}^{v_{max}}   \frac{ e^{- \gamma_{min} q' v'^3 s(t_0)} - e^{- \gamma_{max} q' v'^3 s(t_0)}   }{ q' \, v'^3} \, P(q') P(v') \, dq' \, dv'  }.
\end{eqnarray}

Marginalizing over $\gamma$ gives: 

\begin{eqnarray}
P(q, v | t_0) = \frac{(e^{- \gamma_{min} q v^3 s(t_0)} - e^{- \gamma_{max} q v^3 s(t_0)} ) \, P(q)  \, P(v)}{  q \, v^3 \,  \int_{q_{min}}^{q_{max}} \int_{v_{min}}^{v_{max}}   \frac{ e^{- \gamma_{min} q' v'^3 s(t_0)} - e^{- \gamma_{max} q' v'^3 s(t_0)}   }{ q' \, v'^3} \, P(q') P(v') \, dq' \, dv'  }.
\end{eqnarray}
\end{widetext}

The above SIA posterior depends on the behavior of the quantity $\frac{1}{q \, v^3} (e^{- \gamma_{min} q v^3 s(t_0)} - e^{- \gamma_{max} q v^3 s(t_0)} )$. There are (at least) three cases in which it simplifies greatly, leading to a clear interpretation.

\subsection{The ``conventional SETI" limits}

Suppose our prior assumptions are dominated by a conviction that either $v \rightarrow 0$ or that $q \rightarrow 0$. That is, for whatever reason, we are absolutely sure that the maximum practical speed of expansion is close zero, or that human-stage civilizations almost never embark on cosmic expansion. This would be reflected by a $P(v)$ or a $P(q)$ with support arbitrarily close to zero.

In either limit, the value of $\frac{1}{q \, v^3} (e^{- \gamma_{min} q v^3 s(t_0)} - e^{- \gamma_{max} q v^3 s(t_0)} )$ approaches $ (\gamma_{max} - \gamma_{min}) \, s(t_0) $. And in that case, equation 9 reduces to:

\begin{eqnarray}
P(q, v | t_0) = P(q) \, P(v).
\end{eqnarray}

That is, our anthropic information makes no change to our prior assumptions about $q$ and $v$.\footnote{Conceptual questions can emerge in these limits. What is the interpretation of $q$ in the limit that $v \rightarrow 0$? Presumably, $q$ would be the probability for a human-stage civ to become expansionistic at a subgalactic scale, or even the scale of a single solar system. Effectively, this limit becomes the world-model of 20th century SETI and science fiction.}

This can be regarded as a basic consistency check on our framework. Assuming from the beginning, with complete certainty, that expanding civilizations are irrelevant (being either too slow or too rare to matter) ensures they play no role in observer-selection effects.

\subsection{The regime of expanding civilizations}

If we stay away from the previous limits, so the support of both $P(v)$ and $P(q)$ is away from zero, then we can make another simplification. Suppose we take $\gamma_{min}$ to be tiny, in particular $\gamma_{min} << \frac{1}{s(t_0)}$. And we also take $\gamma_{max}$ to be large, so that $\gamma_{max} >> \frac{1}{q_{min} v_{min}^3 s(t_0)}$. Then we can approximate $(e^{- \gamma_{min} q v^3 s(t_0)} - e^{- \gamma_{max} q v^3 s(t_0)} ) \approx 1$ over the entire domain of interest, and equation 9 reduces to:

\begin{eqnarray}
P( q, v| t_0) = C_{q,v} \, \frac{P(q)}{q} \, \frac{P(v)}{v^3}
\end{eqnarray}
where $C_{q,v} = (\int_{v_{min}}^{v_{max}} \int_{q_{min}}^{q_{max}} \frac{P(q')}{q'} \, \frac{P(v')}{v'^3} dq' dv')^{-1}$ is a normalizing constant.

Marginalizing, this gives:

\begin{eqnarray}
P( q | t_0) = C_q \, \frac{P(q)}{q}
\end{eqnarray}
with $C_q = ( \int_{q_{min}}^{q_{max}} \frac{P(q')}{q'} dq')^{-1}$, and  

\begin{eqnarray}
P( v | t_0) = C_v \, \frac{P(v)}{v^3}
\end{eqnarray}
with $C_v = (\int_{v_{min}}^{v_{max}} \frac{P(v')}{v'^3} dv')^{-1}$ .

Equations 11, 12, and 13 are our first major conclusions. By adopting a world-model in which expanding cosmological civilizations are possible, SIA tells us we should update our priors over $q$ and $v$ according to $P(q) \rightarrow \frac{P(q)}{q} $ and $P(v) \rightarrow \frac{P(v)}{v^3}$, properly normalized.

SIA wants to tilt our assumptions to a lower-than expected probability that human-stage civilizations advance to cosmic expansion (lower $q$), and with an expansion speed slower than we first supposed (lower $v$). Why? \emph{It allows for more observers with anthropic information identical to our own.} Keeping $q$ down (the probability to expand) means the appearance rate of civilizations like ours can be high, while leaving plenty of free space in which to appear. Keeping $v$ low means the universe saturates with ambitious life more slowly, leaving more opportunity for spontaneous appearance of human-stage life here at $t_0$. 

These predictions do not depend explicitly on the details of the background cosmology, or even on our particular appearance time, as $s(t_0)$ falls out of the calculation in equation 11.

This is a consequence of the domain of $\gamma$ we assumed in our approximation, but it is not very restrictive. The requirement of $\gamma_{min} << \frac{1}{s(t_0)}$ means, for our background cosmology, that $\gamma_{min} << 10^{-3}$ (appearances of human-stage civilizations per $Gly^3$ per $Gyr$), which is by no means outside the realm of possibility --- human-stage life could be a cosmic rarity. The requirement of $\gamma_{max} >> \frac{1}{q_{min} v_{min}^3 s(t_0)}$, if $v_{min} \approx 0.1$ and $q_{min} \approx 10^{-4}$, would mean $\gamma_{max} >> 10^4$ (appearances of human-stage civilizations per $Gly^3$ per $Gyr$), which is also quite a pedestrian number\footnote{The Laniakea Supercluster alone contains approximately $10^5$ galaxies, stretched over a distance of about 0.5 $Gly$.}.

\section{Updates based on search results}

SIA has shifted our prior assumptions about $q$ and $v$, but we can go a step further --- suppose we perform an exhaustive galaxy survey, covering fraction $frac$ of the sky (or for any reason expecting to detect fraction $frac$ of ambitious civilizations on our past light cone), looking for expanding civilizations.

If one or more expanding civilizations were detected, we could infer $v$ from the visible geometry of their expanding domains~\cite{olson2016}, and thus get an excellent idea of the practical speed limit we are likely to encounter in the future. But suppose the survey detects $n=0$ such civilizations. This too will update our assumptions about $v$.

Since appearances are a Poisson process (random, independent events), with expected number $E_{\gamma,q,v}=\gamma q (1 - v^3) s(t_0)$ on our past light cone (and $frac \, E_{\gamma,q,v}$ expected to be visible to our survey), the probability to see $n=0$ is given by:
\begin{eqnarray}
p_{\gamma,q,v}(n=0) = e^{- frac \; E_{\gamma,q,v}} = e^{- frac \; \gamma q (1 - v^3) s(t_0)} .
\end{eqnarray}

Now, assuming a survey detects $n=0$ civilization, we can invoke Bayes' theorem to update our previous SIA estimate from equation 8:
\begin{widetext}
\begin{eqnarray}
P(\gamma,q, v | t_0, n=0) &=& \frac{p_{\gamma,q,v}(n=0)  P(\gamma,q, v | t_0)}{\int \int \int p_{\gamma',q',v'}(n=0)  P(\gamma', q',v' | t_0) \, d \gamma' \, d q' \, dv'} \\
& = & \frac{s(t_0) \, e^{- q \gamma s(t_0) ( v^3 + frac (1- v^3))} \, P(q) \, P(v)}{\int \int \frac{ (e^{- q \gamma_{min} s(t_0) ( v^3 + frac (1- v^3))} - e^{- q \gamma_{max} s(t_0) ( v^3 + frac (1- v^3))}  ) \, P(q') \,P(v')}{ q' \,  (  v'^3 + frac (1- v'^3) ) } dq' dv'}.
\end{eqnarray}
\end{widetext}

If we take the same condition on $\gamma$ as that preceding equation 11, and marginalize, we obtain the following estimates:

\begin{eqnarray}
P(q | t_0, n=0) = P(q | t_0) = C_q \, \frac{P(q)}{q} .
\end{eqnarray}

That is, a survey resulting in $n=0$ detections does \emph{not} update our previous SIA-based estimate of $q$. However, it does update our knowledge of $v$:

\begin{eqnarray}
P(v | t_0, n=0) = C_{surv} \frac{P(v)}{ v^3 + frac \; (1 - v^3)}. 
\end{eqnarray}
where $C_{surv} = (\int \frac{P(v')}{ v'^3 + frac \; (1 - v'^3)} dv')^{-1}$. This is our second major result.

Two limiting cases are of immediate interest. In the limit that $frac \rightarrow 0$, i.e. when our survey never had any chance of detecting anything, then we retain our earlier result (equation 13) --- a useless survey changes nothing:

\begin{eqnarray}
\lim_{frac \rightarrow 0} P(v | t_0, n=0) =  C_v \frac{P(v)}{v^3}.
\end{eqnarray}

However, in the limit that $frac \rightarrow 1$, i.e. when our survey was guaranteed to detect every expanding civilization on our past light cone (but still sees none), then we get:

\begin{eqnarray}
\lim_{frac \rightarrow 1} P(v | t_0, n=0) =  P(v).
\end{eqnarray}

The effect of confirming exactly $n=0$ expanding civilizations exist on our past light cone is to perfectly cancel our earlier SIA-induced shift of $\frac{1}{v^3}$. That is, our estimate of $v$ goes back to what it was in the very beginning, before we had invoked any anthropic arguments at all. \emph{An ideal search cancels the SIA effect}, where $v$ is concerned.  

How can seeing $n=0$ civilizations have such a powerful effect? Although we began with independent, uncorrelated prior assumptions about $v$ and $\gamma$, $P(v)$ and $P(\gamma)$, they became highly correlated in the SIA posterior, equation 8. The predicted lower values of $v$ are associated with higher values of the appearance rate, $\gamma$. Thus, the first SIA result (reducing our estimates for $v$) has the effect of greatly amplifying the number that should be visible, on our past light cone. Seeing $n=0$ is then a surprise, and we get a large reversal to our estimate for $v$.

As noted above, if $n > 0$ civilizations are detected, we could infer $v$ directly, but what happens to our knowledge of $q$? In that case, due to the Poisson property, we have:

\begin{eqnarray}
p_{\gamma,q,v}(n) = \frac{(frac \; E_{\gamma,q,v})^n}{n!} e^{- frac \; E_{\gamma,q,v}}.
\end{eqnarray}

Inserting into equation 15 and marginalizing over $\gamma$ yields the same result as before for $q$, reproducing equation 17 (in the high-$\gamma_{max}$, low-$\gamma_{min}$ approximation). Survey results simply do not update $q$. 

\section{Human Existential Risk}

As defined above, $q$ represents the probability for a human-stage civilization to advance to cosmic expansion. Failure to expand could come from many causes, from ``late filter'' extinction events to voluntarily abandonment of expansionist behavior, so it may not be obvious where the additional SIA-induced risk is lurking. Is there a way to interpret the cause as purely benign? Unfortunately, the answer appears to be ``no'' --- this result unavoidably increases all independent sources of existential risk.

\subsection{Refining the risk calculation}

To see the reason for this claim, let us decompose the probability $q$ into $n$ independent probabilities, so that $q = q_1 \, q_2 \dotsb q_n$, with priors $P(q) = P(q_1) \, P(q_2) \dotsb P(q_n) $. That is, the probability to expand is the product of the probability to successfully navigate $n$ universal challenges on the road to expansion. Then, repeating the SIA calculation from section III gives the same update for each probability, i.e. $P(q_1) \rightarrow \frac{P(q_1)}{q_1}$, $P(q_2) \rightarrow \frac{P(q_2)}{q_2}$, $P(q_n) \rightarrow \frac{P(q_n)}{q_n}$, etc. 

The interpretation is that \emph{all independent causes} for a failure to expand are amplified in the same way, provided they are universal risks that all human-stage civilizations must face on the road to technological maturity.

For example, in a two-part decomposition, $q_1$ could represent the probability for a human-stage civilization to survive until the necessary technologies become available, and $q_2$ could represent the probability to expand, given survival. The SIA result tells us to become more pessimistic about both $q_1$ and $q_2$. Thus, we cannot avoid the conclusion that our estimate of humanity's near-term existential risk is amplified.

\subsection{Updating the risk estimate}

In section IV, we noted that survey results do not update the SIA result for $q$. But there is a way to update our estimate of $q$. If humanity were to actually initiate cosmic expansion, it would give us additional information on the true level of risk.

The likelihood function for $q$, given that humanity achieves cosmic expansion, is simply $L(q | \textrm{expand}) = q$. Thus, on the day humanity successfully initiates cosmic expansion (call that day ``E-Day''), we can apply a Bayesian update to the previous SIA estimate of $q$ to get:

\begin{eqnarray}
P(q | t_0, \textrm{expand}) &=& \frac{ L(q | \textrm{expand}) \, \frac{C_q \, P(q)}{q}}{\int_{q_{min}}^{q_{max}} L(q' | \textrm{expand}) \, \frac{C_{q'} \, P(q')}{q'} \, dq'  } \\
&=& P(q).
\end{eqnarray}

Thus, successfully making it to E-Day completely reverses the earlier SIA-induced shift to $q$, taking us back to our original prior estimate $P(q)$. The effect of making it to E-Day is analogous to the effect of a ``perfect survey'' on our estimate of $v$. Unfortunately, this update will then represent our knowledge of $q$ for the \emph{rest of the universe}, coming too late to serve as a guide to humanity's own future.

\section{Consequences and Conclusions}

The SIA re-weighting of prior assumptions can be shockingly powerful. As an illustration, suppose you attended two lectures, back-to-back, by two equally convincing experts --- the first discussing insurmountable difficulties of relativistic space flight, and the second discussing the vast capabilities of a technologically mature civilization. 

Emerging from the lecture hall, you find yourself conflicted, splitting your credence equally between two views of the universe. One in which the maximum practical expansion speed of an ambitious civilization is about $v = .1$, and in the other view somewhere around $ v = .9$. Then, invoking the SIA together with the possibility of expanding civilizations implies you should be $99.9 \%$ convinced of the low-speed scenario (because it leaves room for $\approx 1,000$ times as many human-stage civilizations to have originated at the present time).

The strength of this $\frac{1}{v^3}$ re-weighting implies a prediction. If SIA is the appropriate anthropic school, we expect the maximum practical spacecraft speed to be surprisingly low, given current projects like Breakthrough Starshot~\cite{merali2016}. Instead of a smashing success that implies far-faster spacecraft are possible, our default prediction for Breakthrough Starshot is a set of frustrating, unforeseen difficulties that signal current goals are over-ambitious, and likely close to the maximum practical speed for any advanced civilization (at least when scaled up to extremely long-range, long-duration missions). This may be regarded as a near-term predictive test of the SIA in a cosmological setting.

One important question we have not attempted to answer here is the relevancy of current astronomical data. We have, in fact, looked at the sky and detected $n=0$ expanding cosmological civilizations. But what fraction $frac$ of expanding civilizations on our past light cone should current astronomical data have been expected to notice? Finding an appropriate answer is important to the strength of our $v$-prediction.

The lowering of estimates for $q$ (the probability for human-stage civilizations to advance to cosmic expansion) is troubling for two reasons. First, while a low value of $q$ could have many causes (some benign~\cite{joy2000}), section V shows that all independent, universal sources of existential risk~\cite{ord2020} are unavoidably amplified by our result as well, making late filter extinction~\cite{hanson1998a} more probable than before. Second, unlike estimates of $v$, the results of a sky survey cannot reverse the shift to lower estimates of $q$. The pessimistic estimates are only reversed \emph{after} humanity has successfully navigated any late filters. 

On the other hand, this amplified existential risk follows from the same SIA-based reasoning as our lowered expectations for $v$. This puts particular emphasis on our $v$-predictions. If high values of $v$ appear increasingly practical in the near term as technology improves, discrediting our low-$v$ prediction, the prediction of increased existential risk for humanity would be discredited along with it. In that case, something about our cosmology or use of the SIA would apparently be broken. There do exist a number of reasons to be skeptical of the SIA~\cite{bostrom2003b,shulman2012}, as with every school of anthropic probability, so it is quite possible. However, if we are correct about $v$, with high values looking more impractical with time, the issue of existential risk for humanity becomes ever more urgent. In this sense, our pessimism for human survival is subject to a near-term experimental test.

Finally, we note this is not the first time alien life has been combined with the SIA to adjust our intuition for humanity's survival. Grace noted~\cite{grace2010anthropic, grace2010blog} that the SIA supports a Milky Way Galaxy that is packed with human-stage life (i.e. with anthropic info like ours), but combined with negative SETI results, this implies a late filter awaiting human-stage civilizations. This is based on a different world model than we have used, but it is interesting to note how general such conclusions can be.   

\begin{acknowledgements}
	
We are grateful to Milan {\'C}irkovi{\'c}, for his helpful comments.

\end{acknowledgements}

\bibliography{ref5}{}

\end{document}